\documentclass[12pt]{article}
   \usepackage{putex, epsfig}
 \usepackage{graphicx}
 \usepackage{placeins}

 \newcommand{\beq}[1]{\begin{eqnarray}\label{#1}}
 \newcommand{\eeq}{\end{eqnarray}}

\begin{document}

\institution{ITP-CAS}{Institute of Theoretical Physics, Chinese
Academy of Science}

\title{Jet quenching parameter of Sakai-Sugimoto model}
 \authors{Yi-hong Gao, Wei-shui Xu and Ding-fang Zeng\footnote{gaoyh@itp.ac.cn, wsxu@itp.ac.cn and dfzeng@itp.ac.cn}}

 \abstract{
 Using gauge theory/string duality, we calculated the jet
 quenching parameter $\hat{q}$ of the Sakai-Sugimoto model in various phases.
 Being different from the $\mathcal{N}=4$ SYM
 theory where $\hat{q}\propto T^3$,
 we find that $\hat{q}\propto T^4/T_d$, where
 $T_d$ is the critical temperature of the confined/deconfined
 phase transition. By analyzing the $\hat{q}$ in different phases of
 this theory, we get better understanding about some statements in
 previous works, such as the non-universality
 and the explanation of discrepancies between the theory
 predictions and experiments.
 }

 \date{June 2007}

 \maketitle

 \section{Introduction}\label{introductionSection}

 Experimental relativistic heavy ion collisions produced many
 evidences signalling the quark gluon plasma's formation
 \cite{ex-Adms05, ex-Adcox04} at RHIC.
 One of the characteristic features discovered from data is the
 suppression of heavy quark's spectrum in high-$p_T$
 region \cite{ex-Adler05, ex-Bielcik05}. This is the jet quenching
 phenomenon. Its cause is the medium induced
 heavy quarks' energy loss when they move through QGP.
 So successful models explaining this phenomenon
 usually involve a medium sensitive ``jet quenching parameter''
 $\hat{q}$. While in \cite{jetquench-LRW},  the authors
 proposed a first principle, non-perturbative quantum field
 theoretic definition for it.
  Using their new definition and AdS/CFT correspondence
 \cite{ads-Maldacena,ads-Witten9803,ads-AGMOOreview}, these
 authors calculated the jet quenching parameter of
 the thermal $\mathcal{N}=4$ suppersymmetric Yang-Mills theory.
 After some reasonable parameters's choice adapted to the Au-Au
 collisions at RHIC and the universality assumption
 $\hat{q}_{QCD}\approx\hat{q}_{\mathcal{N}=4}$, they
 find that the theoretically predicted $\hat{q}$ is less
 than that suggested by experiments.
 And the author of \cite{jetquench-Buchel} generalizes the proposal
 of \cite{jetquench-LRW} to the strongly coupled non-conformal
 gauge theory plasma and find that instead of universal, the jet
 quenching parameter is gauge theory specific. They explicitly
 show that the jet quenching parameter increases as one goes
 from a confining gauge theory to a conformal theory. So they
 concluded that the reasons about the jet quenching parameter predicted by \cite{jetquench-LRW}
 being less than the experimental measures is probably due to the existence of some extra energy loss mechanisms of
 the heavy quarks besides gluon radiation.

 In order to investigate whether the jet quenching parameter is universal or not,
 the calculations
 of this parameter in some different gauge theories are meaningful and will help us
 to understand discrepancies between theory and experiments more deeply.
 So after \cite{jetquench-LRW} and
 \cite{jetquench-Buchel}, many other works related to the jet quenching
 parameter appear, see e.g. \cite{jetquench-Portiz},
 \cite{jetquench-CaceresGuijosa},
 \cite{jetquench-LinMatsuo},
 \cite{jetquench-AvarmisSfetsos},
 \cite{jetquench-ArmestoEdelsteinMas} and \cite{jetquench-NTW}\footnote{For more references, one can
 see the paper \cite{Liu:2006he}}.
 One common feature of these works is that, in the dual gravity
 description of the corresponding gauge field theory,
 the background metric usually involves
 an asymptotically $AdS_5$ component.

 An interesting question is,
 how will the jet quenching parameter behave in a gauge theory
 whose dual gravity descriptions involves no $AdS_5$ component?
 In this paper, we will investigate this problem by using the Sakai-Sugimoto
 model \cite{SSmodel}. It is a holographic model of
 the low energy QCD constructed from the intersecting branes. whose
 most striking feature is its phase structure \cite{SSmodel-phase}: (i) the zero and low temperature
 phase or confining phase in which all fundamental field
 degrees of freedom is confined; (ii) in the deconfined phase, there exists the
 chiral symmetry breaking phase and the chiral symmetry restoration phase.

 We will follow the same routine as in \cite{jetquench-LRW},  and calculate the jet quenching parameter.
 After doing some analysis this parameter in the different phase of this model, we
 can get more deep understanding about the jet quenching
 phenomenon in the thermal QCD-like theory\footnote{In \cite{drag-Talavera, SSmodel-meson},
 the authors have studied the
 drag force and screening length problems in this holographic
 model.}.

 The organization of this paper is as follows,
 we will briefly review the Sakai-Sugimoto model in the next section,
 then calculate the jet-quenching parameter of
 this model in different phases in the section 3, while
 the last section contains the main conclusions of the paper.

 \section{Brief review of Sakai-Sugimoto model}\label{braneConstructSection}

 In the \cite{SSmodel}, Sakai and Sugimoto constructed a holograhic QCD-like model from the
 brane components D4, D8 and $\overline{D8}$ in type
 IIA string theory. In this brane construction, the low
 energy effective theory in the intersecting dimension is a $U(N_c)$
 gauge theory with $N_f$ flavors. And there also exists a global $U(N_f)_L\times
 U(N_f)_R$ chiral symmetry.

  In the strong coupling region, we can use the supergravity
approximation to analyze the
 underlying physics. Thus, in the zero-temperature phase, we can use the D8 brane to probe
 the following background
 \beq{}
 &&\hspace{-5mm}ds^2_c=H^{-\frac{1}{2}}(-dt^2+d\vec{x}\cdot
 d\vec{x}+fdx_4^2)+H^{\frac{1}{2}}(f^{-1}du^2+u^2d\Omega_4^2),\nonumber\\
 &&\hspace{-5mm}H=\frac{R_{D4}^3}{u^3},\ f=1-\frac{u_\Lambda^3}{u^3}, ~R_{D4}^3=N_c \pi g_s l_s^3,\nonumber\\
 &&\hspace{-5mm}e^\phi=g_s\left[\frac{u^3}{R_4^3}\right]^{1/4},\
 F_{(4)}=\frac{2\pi N_c}{V_4}\epsilon_4,
 \label{zeroTempBGRND}
 \eeq
 where $t$, $\vec{x}$ are four uncompactified world-volume
 coordinates of the D4 branes,  and $x_4$ denotes the compactified
 world-volume coordinate. The sub-manifold spanned by $x_4$
 and $u$ is singular as $u\rightarrow u_\Lambda$, to avoid this
 singularity, the periodical identification of $x_4$ should be
 \beq{}
 x_4\sim x_4+2\pi R,\ 2\pi
 R=\frac{4\pi}{3}\left[\frac{R_{D4}^3}{u_\Lambda}\right]^{1/2}.
 \eeq

  While to the high temperature phase of the gauge theory i.e $T>(2\pi R)^{-1}\equiv
  T_d$ (where $T_d$ is the critical temperature of the Hawking-Page phase
  transition),
 the dual gravity system's background becomes
 \beq{}
 &&\hspace{-5mm}ds^2_d=H^{-\frac{1}{2}}(-fdt^2+d\vec{x}\cdot
 d\vec{x}+dx_4^2)
 +H^{\frac{1}{2}}(f^{-1}du^2+u^2d\Omega_4^2)\nonumber\\
  &&\hspace{-5mm}H=\frac{R_{D4}^3}{u^3},\ f=1-\frac{u_T^3}{u^3}.
 \label{highTempBGRND}
 \eeq
 The others are same as in (\ref{zeroTempBGRND}), and here the Hawking temperature
 is $T=\frac{3}{4\pi}\left[\frac{u_T}{R_{D4}^3}\right]^{1/2}$.

 As investigated in \cite{SSmodel-phase}, there exists two phases in this deconfined phase.
 We can use the $D8$- and
 $\overline{D8}$-branes to probe the background (\ref{highTempBGRND}).
 The world-volume coordinates of the $D8$- and $\overline{D8}$-branes
 consists of $t$, $\vec{x}$, $u$ and $\Omega_4$.
 Assuming that as $u\rightarrow\infty$ the D8 branes
 sit at $x_4=0$ while the $\overline{D8}$-branes
 at $x_4=D$, then the induced metric
 on the probe $D8$- or $\overline{D8}$-branes reads
 \beq{}
 ds^2_8=H^{-\frac{1}{2}}(-fdt^2+d\vec{x}\cdot
 d\vec{x})+(H^{-\frac{1}{2}}x_4^{\prime2}
 +H^{\frac{1}{2}}f^{-1})du^2+H^{\frac{1}{2}}u^2d\Omega_4^2.
 \label{highTempInduceBGRND}
 \eeq
 The DBI action of the probe D8-branes gives equation of motion,
 \beq{}
 &&\hspace{-5mm}\frac{u^4f}{\sqrt{f+H/[\frac{dx_4}{du}]^2}}=
 u_0^4\sqrt{f(u_0)}\nonumber\\
 &&\hspace{-5mm}\Rightarrow\frac{dx_4}{du}=
 \pm\left[\frac{H}{f([u^8f]/[u_0^8f_0]-1)}\right]^{1/2}
  \label{D8profile}\eeq
 where $u_0$ is the $u$-coordinate of a point at which
 the $x_4^\prime(u)=0$. As the result,
 \beq{}
 x_4(u)=\frac{D}{2}\pm\left[\frac{R_4^3}{u_0}\right]^{1/2}\int_1^{\frac{u}{u_T}} dx
 \left[\frac{x^{-3}}{(1-y_T^3x^{-3})[x^8(1-y_T^3x^{-3})/(1-y_T^3)-1]}\right]^{1/2}
 \label{chisbD8branes}
 \eeq
 where $y_T=\frac{u_T}{u_0}$. In this case, the $D8$- and
 $\overline{D8}$ are connected smoothly at $x_4=D/2$ point.

\begin{figure}[ht]
 \begin{center}
 \centering\includegraphics[]{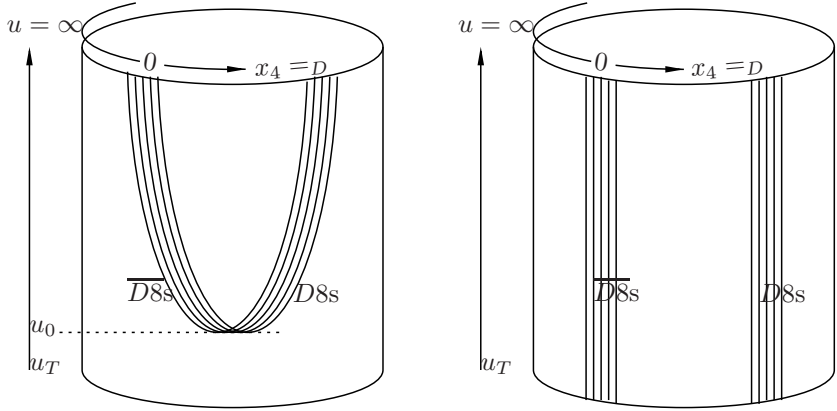}
 \end{center}
 \caption{
 The chiral symmetry is broken in the left one, however, it is
 restored in the right one.
 }
 \label{DDbarProfile}
 \end{figure}

 However, the configuration constrained by
 \beq{}
 \partial_ux_4\equiv0
 \eeq
 also satisfy eq(\ref{D8profile}).  This means that the $D8$ and $\overline{D8}$
 are located at fixed positions
 \beq{}
 x_4(u)\equiv0~
 \textrm{\ or\ }~
 x_4(u)\equiv D.
 \label{chisrD8branes}
 \eeq
 Between the phases (\ref{chisbD8branes}) and (\ref{chisrD8branes}), as pointed out by \cite{SSmodel-phase},
 there exists a first order chiral phase
 transition. Under a critical temperature
 $T_{\chi SB}\simeq 0.154/D$, the chiral symmetry is broken. However, above
 this temperature, the symmetry will be restored.
 We depict the relevant profile in the left hand side of Figure \ref{DDbarProfile}.

 \section{The jet quenching parameter}\label{eomSection}

 \subsection{General definitions}\label{gRoutineSection}

 By the definition of \cite{jetquench-LRW}, the jet quenching
 parameter $\hat{q}$ is related to the expectation value of a
 light-like wilson loop $<W^A(\mathcal{C})>$ in the adjoint
 representation whose contour $\mathcal{C}$ composed of
 a rectangle of length $L^-$ and width $L$ with $L<<L^-$ and
 $L<<T^{-1}$. Motivated by
 dipole approximation \cite{dipoleApprox}
 \beq{}
 <W^A(\mathcal{C})>\approx\exp\left[-\frac{1}{4\sqrt{2}}\hat{q}L^-L^2\right].
 \label{jetQuenchPDef}
 \eeq
 The author of \cite{jetquench-LRW} defined $\hat{q}$ as the
 coefficient of the $L^-L^2/(4\sqrt{2})$ term in
 $\ln<W^A(\mathcal{C})>$. So
 \beq{}
 \hat{q}=\frac{-4\sqrt{2}}{L_-L^2}\ln<W^A(\mathcal{C})>.
 \label{jetQuenchFinalDefinition}
 \eeq
 The factor of $\sqrt{2}$ in the above equations comes from the
 light-cone coordinate definition $x^{\pm}=(t\pm x^1)/\sqrt{2}$. If
 a different definition $x^{\pm}=t\pm x^1$ are used,
 the factor of $\sqrt{2}$ in the equation (\ref{jetQuenchPDef}) will not
 appear, see for e.g. \cite{jetquench-Buchel}.

 So, let us consider such a Wilson loop following the same routine
 of \cite{jetquench-LRW}. In the planar limit, wilson loops
 in the adjoint representation is related
 to that in the fundamental representation through
 \beq{}
 \ln<W^A(\mathcal{C})>=2\ln<W^F(\mathcal{C})>.
 \eeq
 According to the gauge theory/string duality, see for e.g.
 \cite{wilson-Maldacena}, \cite{wilson-ReyYee},
 \cite{wilson-ReyTheisenYee} and \cite{wilson-BISY},
 $<W^F(\mathcal{C})>$ can be given by
 \beq{}
 <W^F(\mathcal{C})>=\exp[i\hat{S}(\mathcal{C})]
 \eeq
 where $\hat{S}$ is related to the
 extremal action $S$ of a fundamental string whose
 world-sheet on the boundary of the background space-time have
 $\mathcal{C}$ as the boundary, i.e.
 \beq{}
 &&\hspace{-5mm}S=-\frac{1}{2\pi\alpha'}\int d^2\sigma\sqrt{-\gamma},\nonumber\\
 &&\hspace{5mm}\gamma=\textrm{det}\gamma_{ab},
 \ \gamma_{ab}=\partial_aX^\mu\partial_bX^\nu G_{\mu\nu}
 \label{Sdefinition}
 \eeq
 which is just the Nambu-Goto action of the string.
 In the above equations, $X^\mu$ is the embedding of the
 fundamental string in the background space-time with metric
 $G_{\mu\nu}$. From gauge theory, since the action $S$ contains the
 self-energy $S_{sef}$
 of quarks attached on the
 two ends of the string, it is usually divergent. In order to describe the interactions
 between the two quarks, we need subtract the self-energy part. The
 final part $\hat{S}$ will be
 \beq{}
 \hat{S}=S-S_{sef}
 \eeq

 \subsection{String configuration}\label{strConfigSection}

  According to the general routines in the last section,
 in order to calculate the jet quenching parameters of the Sakai-Sugimoto
 model in different phases, we can write the background space-time (\ref{zeroTempBGRND}) and
(\ref{highTempBGRND}) in the light-cone
 coordinates. The zero temperature background becomes
 \beq{}
  ds^2=
 H^{-\frac{1}{2}}(-2dx^+dx^- +dx_\bot^2+fdx_4^2)+H^{\frac{1}{2}}(f^{-1}du^2+u^2d\Omega_4^2)
    \label{lightConeMetric1}
 \eeq and the high temperature one is\beq{}
 ds^2&=&H^{-\frac{1}{2}}(-\frac{1+f}{2}dx^+dx^- +
 \frac{1-f}{4}[dx^{+2}+dx^{-2}]+dx_\bot^2+dx_4^2)\nonumber\\
 &&~~~~~+H^{\frac{1}{2}}(f^{-1}du^2+u^2d\Omega_4^2).\label{lightConeMetric}\eeq{}

 To find the embedding of strings into the above backgrounds, we give the following ansatz for the string
 configuration,
 \beq{}
  X^-=\tau, ~X^2=\sigma, \nonumber \\ u(\sigma),~~~ X^4(\sigma)
 \eeq
 Under this ansatz, the Nambu-Goto action
 of the string vanishes in the zero temperature metric (\ref{lightConeMetric1}),
 and in the high temperature background (\ref{lightConeMetric}) reads
 \beq{}
S=-\frac{1}{2\pi\alpha^\prime}\int d^2\sigma
   \sqrt{\frac{f-1}{2}\left(\frac{1}{H}
        +\frac{u^{\prime2}}{f}
        +\frac{u^{\prime2}}{H}\left[\frac{dx_4}{du}\right]^2\right)}
 \label{NambuGotoAction}
 \eeq{}
 where $u^\prime$ denotes $\frac{du}{d\sigma}$.

 By minimizing the Nambu-Goto action, we can find the equation of motion and then the
 solution of the required string configuration. Obviously, since the action
 equals to zero identically in the zero-temperature phase, the
 corresponding string configuration satisfying required boundary
 conditions cannot be determined uniquely. Notice that it
 is $\hat{S}$ instead of $S$ that determines the jet quenching
 parameter. Hence this means the jet quenching parameter $\hat{q}$ will be equal to zero.
 Since the jet quenching parameter
 is designed to describe the property of medium through which
 the test heavy quark moves. The vanishing jet quenching parameter $\hat{q}$ means that
 the medium will look transparent to the test heavy quarks in the confining phase.

 For the high temperature phase, we can get the equation of motion from the action (\ref{NambuGotoAction}) which reads
 \beq{}
 u^{\prime2}=\frac{c^2f}{H+(dx_4/du)^2f}
 \hspace{5mm} \label{eomFirstCHISB}
 \eeq
 In deriving these equations, we used the fact that
 Lagrangian $\mathcal{L}\equiv\sqrt{-\gamma}$ in
 the equation (\ref{NambuGotoAction}) does not contain $\sigma$ explicitly.
 So the effective ``hamiltonian''
 $\mathcal{H}\equiv u^\prime\frac{\partial\mathcal{L}}{\partial
 u^\prime}-\mathcal{L}$ is a conserved quantity.
 The integral constant $c $ characterizes the profile of string. If
 the $dx_4/du=0$, then the equation (\ref{eomFirstCHISB}) will
 reduce to \beq{}
 u^{\prime2}=\frac{c^2f}{H}
 \hspace{5mm} \label{eomFirstCHISBR}
 \eeq
 due to the fact that $dx_4/du=0$. In this case, the test string
 lies in the quark gluon plasma with chiral symmetry restored.

\begin{figure}[ht]
 \begin{center}
 \centering\includegraphics[]{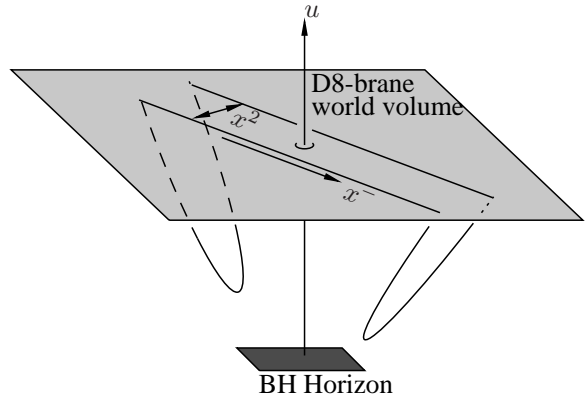}
 \end{center}
 \caption{
 The configuration of strings with light-like boundaries on the
 $D8$-branes world-volume.
 }
 \label{stringConfig}
 \end{figure}

 We require the string has the profile depicted in figure
 \ref{stringConfig} intuitively. 
 The end points of this string are
 located on the $D8$-branes world-volume. And its projection on this worldvolume will form a
 light-like Wilson loop. It goes from the world volume of
 the $D8$-branes at $u=\infty$, towards the black hole horizon
 then turn around at some lowest point and come back to the $D8$-branes
 world-volume again, i.e. the string has a U-shape in
 the $x_2-u$ plane.

 From equation (\ref{eomFirstCHISBR}), we find that in the chiral symmetry restored phase the only zero point
 occurs at the black hole horizon. It means the turning point of the string will exactly reach the
 horizon. And in this phase, the length of the string is
\beq{}&&\hspace{-5mm}L=2\int du\frac{1}{u^\prime}=
 \frac{u_T}{c}\left[\frac{R_{D4}^3}{u_T^3}\right]^{1/2}\frac{2\sqrt{\pi}}{3}\frac{\Gamma[1/6]}{\Gamma[2/3]}\eeq{}
 where we have used the integral \beq{}
 \int_1^\infty\frac{dx}{\sqrt{x^3-1}}
 =\frac{\sqrt{\pi}}{3}\frac{\Gamma[1/6]}{\Gamma[2/3]}.
 \eeq

\begin{figure}[ht]
 \centering\includegraphics[width=0.45\textwidth]{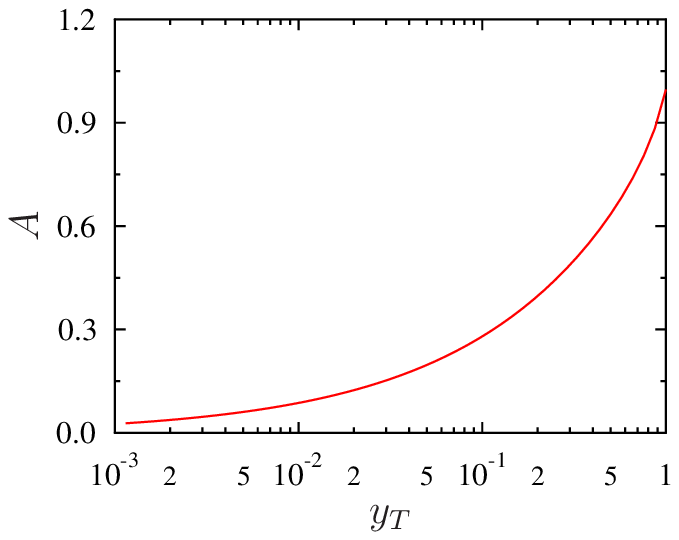}
\includegraphics[width=0.45\textwidth]{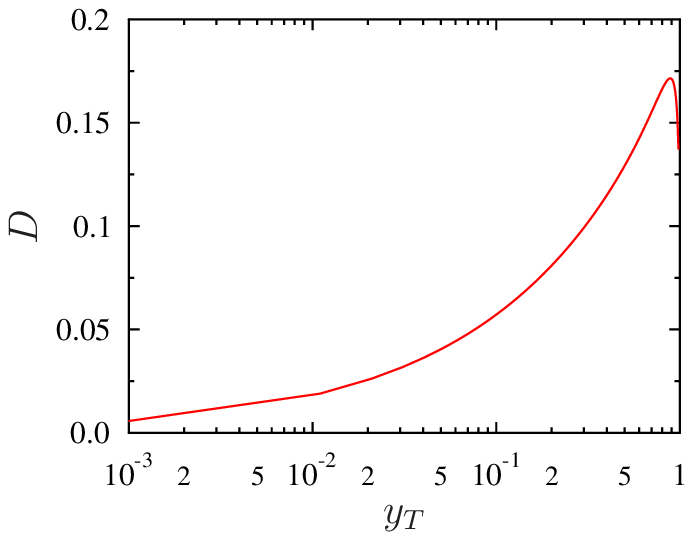}
 \caption{
 The left panel, the numerical behavior of $A[y_T]$ defined in
 the equation (\ref{Adefinition}). The right panel, the distance $D$ between
 the $D8$- and $\overline{D8}$-branes v.s. $y_T$. The distance $D$
 is measured by the temperature inverse, $T^{-1}$. When $y_T<<1$,
 $D\propto y_T^{1/2}$ which can be looked out from
 the equation (\ref{chisbD8branes}).
 }
 \label{AD-yT-numerics}
 \end{figure}

 While in the broken chiral symmetry phase, equation (\ref{eomFirstCHISB}) has
 two zero points, one at the black hole horizon $u=u_T$,
 the other at $u=u_0>u_T$. To assure the positivity of
 $u^{\prime2}$, the lowest point of the string can only reach
 $u=u_0$. Then, the length $L'$ of string is
 \beq{}
 L'&=&\frac{2}{c}\int\left[\frac{H+(dx_4/du)^2f}{f}\right]^{1/2}\nonumber\\
  &=&\frac{u_T}{c}\left[\frac{R_{D4}^3}{u_T^3}\right]^{1/2}
 \frac{2\sqrt{\pi}}{3}\frac{\Gamma[1/6]}{\Gamma[2/3]}\cdot A[y_T]
    \label{Lvsc}
 \eeq
with the definition
 \beq{}
 \int_{u_0/u_T}^\infty\frac{dx}{\left[x^3-1-u_0^8f_0/u_T^8x^{-5}\right]^{1/2}}
 =A[y_T]\cdot\frac{\sqrt{\pi}}{3}\frac{\Gamma[1/6]}{\Gamma[2/3]}
 \label{Adefinition}
 \eeq
 where the $y_T=u_T/u_0$. Obviously, if $A[y_T]=1$, then the length $L'$ of the string in the broken chiral
 symmetry phase will reduce to the corresponding one in the symmetry restored phase. The numerical
 behavior of $A[y_T]$ is displayed in the left hand side of Figure \ref{AD-yT-numerics}.
From the figure we can see the behavior of the function $A[y_T]$
with the variable $y_T$. When $y_T=1$, $A[y_T]$ will reach the
maximal value $1$. This means that at this point the chiral symmetry
broken phase will turn into the chiral symmetry restored phase.

 \subsection{The jet quenching parameter in various
 phases}\label{jetQuenchSection}

 Now we can use the routines of subsection \ref{gRoutineSection}
 to calculate the jet quenching parameters of the Sakai-Sugimoto model.
 We start with the Nambu-Goto action,
 \beq{}
 \hspace{-5mm}S=\frac{L_-\cdot L'}{2\sqrt{2}\pi\alpha^\prime}
  \left[\frac{u_T^3}{R_{D4}^3}\right]^{1/2}\cdot\sqrt{1+c^2}
  \label{NambuGotoActionIntegrated}
 \eeq
 and  the self-energy
 of the string is
 \beq{}
 \hspace{-5mm}S_{sef}=\frac{2}{2\pi\alpha^\prime}\int d\tau
du\sqrt{-G_{--}G_{uu}}
 \nonumber\\
\hspace{6mm}=\frac{L_-u_T}{\sqrt{2}\pi\alpha^\prime}
 \frac{\sqrt{\pi}}{3}\frac{\Gamma[1/6]}{\Gamma[2/3]}\cdot A[y_T]
   \eeq
 Then according to the equation (\ref{jetQuenchFinalDefinition}),
 the jet quenching parameter can be expressed as
 \beq{}
 \hat{q}=2(S-S_{sef})\cdot\frac{4\sqrt{2}}{L_-L'^2}
 =\frac{4}{\pi\alpha^\prime}
 \left(\left[\frac{u_T^3}{R_{D4}^3}\right]^{1/2}\frac{(1+c^2)^\frac{1}{2}}{L'}-
 \frac{u_T}{L'^2}\frac{2\sqrt{\pi}}{3}\frac{\Gamma[1/6]}{\Gamma[2/3]}\cdot
 A[y_T]\right)
 \label{jetQuenchPrimary}
 \eeq
 The right hand side of this equation involves the undetermined
 integration constant $c$. But it can be
 eliminated by the relations expressed in the equation (\ref{Lvsc}). And since the wilson loop satisfies the condition
 $L' T\rightarrow 0$, we find
 \beq{}
 &&\hspace{-5mm}\hat{q}=
  \frac{64\sqrt{\pi^3}}{27}\frac{\Gamma[2/3]}{\Gamma[1/6]}\frac{\lambda}{A[y_T]}T^4/T_d
 \label{jetQuenchAnaResults}
 \eeq
 where we have used the relation $\lambda=g_{YM}^2N_c$, $g^2_{YM}=2\pi g_sl_s/R$,
 and the critical temperature $T_{d}=1/(2\pi R)$. If choosing the
 parameter $A[y_T]=1$, then the above results will reduce to the
 corresponding ones in the chiral symmetry restoration phase.
 However, if $A[y_T]\neq1$, all the results in the above are in the
 chiral symmetry broken phase.

 Let us make some comments on these results in the following.
 First, we compare the jet quenching parameters in the chiral
 symmetry broken phase and that in the chiral symmetry restored
 one. Since $A[y_T]<1$, we know that the jet quenching parameter $\hat{q}$ in the chiral
 symmetry broken phase is larger than the one in the chiral symmetry restored phase. As is
 known, in the chiral symmetry broken phase, the gauge system is a mixture of quark-gluon plasma
 and hadrons; while in the chiral symmetry restored phase, the gauge system
 is a pure quark-gluon plasma with all hadrons resolved. So
 if the dominant mechanism of heavy quarks energy loss is due to
 gluon radiation induced by the medium, we expect the jet quenching
 parameter in the chiral
 symmetry broken  phase be less than that in the chiral symmetry
 restored one, since the gluons density in the former is less
 than that in the latter. So we think
 that our result about the parameter $\hat{q}$ being more large in the chiral symmetry broken phase
 may be looked as an evidence supporting the viewpoint of
 \cite{jetquench-Buchel} which states that, the reasons of the jet
 quenching parameter found in \cite{jetquench-LRW} being less than the experimental measures are
 due to existence of some extra
 energy loss mechanisms(such as binary collisions) of the heavy quarks
 as they move through the mediums. \cite{energyLoss:dominantMechanism}
 even suggest that the dominant mechanism is not gluon radiation
 but binary collisions.

 Second, let us make a comparison between the jet quenching
 parameter of Sakai-Sugimoto model plasmas with the $N=4$ SYM theory
 plasmas. The most striking difference is their
 temperature dependence. The former, $\hat{q}_{SS}\propto T^4$ at a
 fixed $T_d$
 while the latter, $\hat{q}_{\mathcal{N}=4SYM}\propto T^3$. This
 is an evidence that the jet quenching parameter in the
 different gauge theories is not universal. If experiments can
 measure the jet quenching parameter v.s. temperature relation
 explicitly, this may be a good starting point for further exploration of the meaning of
 the jet quenching parameter.

 Finally, we make a little numerics to see whether the
 Sakai-Sugimoto theory plasma fit experiments more better or
 even worse than $N=4$ SYM plasmas.
 For this purpose, we rewrite the equation (\ref{jetQuenchAnaResults}) as
 \beq{}
  \hat{q}=\frac{3.21}{A[y_T]}{\lambda T^4/T_d}
  \label{jetQuenchNumResults}
 \eeq{} If letting $A[y_T]=1$, it reduces to
\beq{} \hat{q}=3.21\lambda T^4/T_d
 \eeq{} in the chiral symmetry restoration phase.

 Taking $N_c=3$, $g_{YM}^2=0.5$ and $T_d=300$
 Mev,
 we find, in order to obtain $\hat{q}=5 \textrm{Gev}^2/\textrm{fm}$,
 the corresponding temperature of the quark gluon plasma in the chiral symmetry restoration phase is
 $280$ Mev, which is smaller than expected \cite{ex-KolbHeinzreview}.
 Equivalently, the jet
 quenching parameter on the basis of these parameters
 are less than that suggested by RHIC data, note that,
 $T\sim\left[\frac{\hat{q}T_d}{N_cg_{YM}^2\hat{q}}\right]^{1/4}$.
 However, in the
 chiral symmetry broken phase, the corresponding temperature is
 $280\cdot A^{\frac{1}{4}}[y_T]$ Mev. As long
 as we choose a appropriate $A[y_T]$, as a result, the jet
 quenching parameter can be equal to the one suggested by RHIC
 data.

 To understand what the smallness
 of $A[y_T]$ means. Let us turn back to the Sakai-Sugimoto model
 for a while. From figure \ref{AD-yT-numerics} we know that smaller $A[y_T]$
 corresponds to smaller $y_T$. From the equation (\ref{chisbD8branes}) we know that $y_T$ is
 related with the distance $D$ between the $D8$- and
 $\overline{D8}$-branes. We numerically integrated this equation
 and plot the $D$ v.s. $y_T$ relation in the right panel of Figure
 \ref{AD-yT-numerics}. From the figure we see that in the small
 $y_T$ region, $D$ is an increasing function of $y_T$. So the
 smallness of $y_T$ means the small distance between the $D8$- and
 $\overline{D8}$-branes. Hence from aspects of the microscopic
 construction of Sakai-Sugimoto model, to enhancing the jet
 quenching parameter, we only need to make the $D8$- and
 $\overline{D8}$-branes sit as near as possible.

 \section{Summary}\label{summarySection}

 In this paper, we calculated the jet quenching parameters of the Sakai-Sugimoto
 model in various phases. In this holographic model, the gravity metric is different from the
 asymptotic $AdS_5$ spacetimes. After some analysis, we find that the jet quenching parameter
  $\hat{q}$ in this
holographic model is proportional to the fourth power of the
 temperature at a fixed confining/deconfining temperature $T_d$, $\hat{q}\propto T^4/T_d$.
 By comparing the jet quenching parameter of this
 model in different phases with that of the $N=4$ SYM plasmas,
 we get more deep understanding about the
 following statements in previous works, (i) the jet
 quenching parameters in different gauge theories are not
 universal, bur are specific for a given gauge theory; (ii) the discrepancies about the jet quenching
 parameter between
 theoretically calculating
 and experimental data is probably due to some extra energy
 loss mechanisms except gluon radiations of the heavy quarks
 when they move through the quark gluon plasmas.

\end{document}